\documentclass[10pt]{article}

\usepackage{amsmath}

\begin{document}


\begin{center}
{\Large {\bf Extension maps}}
\end{center}

\hfil\break

\begin{center}
{\bf Aik-meng Kuah\footnote{kuah@physics.utexas.edu}, E.C.G. Sudarshan}\break
{\it Department of Physics \\
University of Texas at Austin \\
Austin, Texas 78712-1081}\\
\end{center}

\begin{center}
Mar 12 2005
\end{center}

\begin{abstract}

We define extension maps as maps that extend a system (through adding ancillary systems) without changing the state in the original system.  We show, using extension maps, why a completely positive operation on an initially entangled system results in a non positive mapping of a subsystem.  We also show that any trace preserving map, either positive or negative, can be decomposed in terms of an extension map and a completely positive map.  

\end{abstract}

\hfil\break
\hfil\break

\pagebreak


\section{Introduction}

This paper ties in with previous papers on non completely positive maps ~\cite{SudarshanShaji03} and the dynamical evolution of initially open entangled systems ~\cite{JordanShajiSudarshan04}.

It is well known ~\cite{Sudarshan61} that any completely positive linear map of a quantum system can be written as a unitary operation of an extended system where the original quantum state is coupled separably with an ancillary system (subscript $e$) in a known state:

\begin{equation}
\Lambda \rho = Tr_e [u \rho \otimes |e_0\rangle\langle e_0| u^\dagger]
\end{equation}

However it was shown in a recent paper ~\cite{JordanShajiSudarshan04} that dynamics of initially entangled systems need not be given by completely positive maps.  Here we will outline the problem more concisely.

We wish to consider the linear transformation of a subspace of a system that undergoes a unitary evolution:

\begin{equation}
\Lambda \rho = Tr_e [u E(\rho) u^\dagger]
\end{equation}

$E$ is what we call an extension map.  An extension map should satisfy:

\begin{equation}
Tr_e [E(\rho)] = \rho
\end{equation}

This construction of using an extension map simplifies the problem - the map $\Lambda$ is completely positive if and only if the extension map $E$ is completely positive, since the unitary transformation $u$ is by definition completely positive.

For example, to write a completely positive map as a larger unitary transformation ~\cite{Sudarshan61}, the extension map is chosen to be:

\begin{equation}
E(\rho) = \rho \otimes |0_e\rangle\langle 0_e| 
\end{equation}

For the problem outlined in ~\cite{JordanShajiSudarshan04}, we wish to understand the mapping of a subsystem of a larger system, which is initially in a pure entangled state and undergoes unitary evolution.  We find that the extension map has to be either non-linear (ie. like purification ~\cite{purification}), or if linear, can only be positive on certain density matrices.  To demonstrate, consider any pure entangled state $|\Phi\rangle$, the density matrix of a subsystem would be mixed.  For example, in a $2 \otimes 2$ system, we could have:

\begin{equation}
E(\lambda_1 |\phi_1\rangle\langle\phi_1| + \lambda_2 |\phi_2\rangle\langle\phi_2|) = |\Phi\rangle\langle\Phi|
\end{equation}

If the extension map is linear, then we can expand:

\begin{equation}
\lambda_1 E(|\phi_1\rangle\langle\phi_1|) + \lambda_2 E(|\phi_2\rangle\langle\phi_2|) = |\Phi\rangle\langle\Phi|
\end{equation}

The RHS is a pure state, which is convexly non-decomposable.  Therefore $E$ cannot be positive on all states, in this example it cannot be positive on $|\phi_1\rangle$ and/or $|\phi_2\rangle$.

This explains why a non-positive map results from what should be a completely positive operation (unitary transformation); it stems from the choice of a non-positive extension map.

\section{Extension maps for non positive maps}

An interesting question would be to ask: given a non positive linear map $\Lambda$, can a corresponding unitary evolution and extension map be found?  In this section we construct a solution to this problem.

Let us consider only trace preserving maps:

\begin{equation}
Tr [\Lambda \rho] = \sum_{rs}^N \sum_{r'} \Lambda_{r'r,r's} \rho_{rs} = Tr[\rho]
\end{equation}

\begin{equation}
\sum_{r'} \Lambda_{r'r,r's} = \delta_{rs}
\end{equation}

We can write the map $\Lambda$ as the difference of two completely positive hermitian maps.  Let us write the $\Lambda$ in its canonical decomposition:

\begin{equation}
\Lambda = \sum_{i} \lambda_i L_i \times L_i^\dagger
\end{equation}

We simply group the positive eigenvalues/matrices and the negative eigenvalues/matrices to define:

\begin{equation}
\Lambda^{(+)} = \sum_{i | \lambda_i>0} \lambda_i L_i \times L_i^\dagger
\end{equation}

\begin{equation}
\Lambda^{(-)} = \sum_{i | \lambda_i<0} |\lambda_i| L_i \times L_i^\dagger
\end{equation}

Note that by definition, both $\Lambda^{(+)}$ and $\Lambda^{(-)}$ are completely positive.  And we have:

\begin{equation}
\Lambda = \Lambda^{(+)} - \Lambda^{(-)}
\end{equation}

Let us define matrices $J$ and $K$ as follows (this construction was first used in ~\cite{SudarshanShaji03}):

\begin{equation}
J_{sr} = \sum_{r'} \Lambda^{(+)}_{r'r,r's}
\end{equation}

\begin{equation}
K_{sr} = \sum_{r'} \Lambda^{(-)}_{r'r,r's}
\end{equation}

First we note that the matrix $J$ (and similarly $K$) is hermitian:

\begin{equation}
J_{sr} = \sum_{r'} \sum_{i | \lambda_i>0} \lambda_i \{L_i\}_{r'r} \{L_i\}^\dagger_{sr'}
= J_{rs}^*
\end{equation}

Next we note that $J$ cannot be singular if the map $\Lambda$ is trace preserving, since the trace preserving condition is simply:

\begin{equation}
J - K = 1; J>0, K \geq 0
\end{equation}

$J$ and $K$ are partially traced matrices of $\Lambda^{(+)}$ and $\Lambda^{(-)}$ which are completely positive, therefore $J$ and $J$ must be positive.  Given $J = 1 + K$ and the matrices are all positive, it follows that $J$ cannot be singular.

However it is possible that $K$ is singular.  Let us write down the canonical decomposition of $K$ (keeping in mind $K$ is hermitian):

\begin{equation}
K = \sum_q k_q |\psi_q \rangle \langle \psi_q|; k_q>0
\end{equation}

We can define a pseudo-inverse of matrix $K$ as:

\begin{equation}
K^{-1} = \sum_q k_q^{-1} |\psi_q \rangle \langle \psi_q|
\end{equation}

\begin{equation}
K^{-1} K = \sum_q |\psi_q \rangle \langle \psi_q| \equiv \Psi
\end{equation}

Let us also define an orthornormal set of eigenvectors $|\phi_q\rangle$ spanning the singular subspace of $K$:

\begin{equation}
K |\phi_q\rangle = 0
\end{equation}

We can show that $\Lambda^{(-)}$ must destroy all information in this subspace.  Let us consider:

\begin{equation}
Tr[\Lambda^{(-)} ( |\phi_q\rangle\langle \phi_q|)] = 
\sum_{rs} K_{sr} \{\phi_q\}_r \{\phi_q\}_s^* = 0
\end{equation}

Since $\Lambda^{(-)}$ is completely positive and hermitian, and $|\phi_q\rangle\langle \phi_q|$ is positive, if the trace of the result is zero then the result itself must be zero.

Next let us show:

\begin{equation}
\Lambda^{(-)} ( |\phi_q\rangle\langle \psi_r| ) = \Lambda^{(-)} ( |\psi_r\rangle\langle \phi_q| ) = 0
\end{equation}

Let us define the matrices:

\begin{equation}
\begin{array}{c}
A_{mn}=\Lambda^{(-)}_{m\phi_q,n\phi_q} \\
B_{mn}=\Lambda^{(-)}_{m\phi_q,n\psi_r} \\
D_{mn}=\Lambda^{(-)}_{m\psi_r,n\psi_r}
\end{array}
\end{equation}

and the matrix:

\begin{equation}
Z = \left[\begin{array}{cc}
A & B\\
B^\dagger & D\end{array}\right]
\end{equation}

$Z$ is a submatrix of $\Lambda^{(-)}$ so it is non-negative
.  We showed that $A = 0$ in equation 21, therefore it follows $B = B^\dagger = 0$ otherwise $Z$ would be negative.

This gives us a very useful result:

\begin{equation}
\Lambda^{(-)} ( \rho ) = \Lambda^{(-)} ( \Psi \rho) 
\end{equation}

which we obtain by inserting the identity before $\rho$:

\begin{equation}
1 = \sum_u |\psi_u\rangle\langle \psi_u| + \sum_q |\phi_q\rangle\langle \phi_q| 
\end{equation}

Now we can move on to the main result -- defining an extension map and unitary evolution for a non completely positive map $\Lambda$.  Let us define the extension map:

\begin{equation}
E(\rho) = J \rho \otimes |e_{(+)}\rangle\langle e_{(+)}| - K \rho \otimes |e_{(-)}\rangle\langle e_{(-)}| 
\end{equation}

This satisfies the condition:

\begin{equation}
Tr_e[E(\rho)] = (J - K) \rho = \rho
\end{equation}

We note that the dimension of the space needed for this extended state is $dim(J) + dim(K)$, and $dim(J)=N$ since we have noted that $J$ cannot be singular. 

Then let us define a map $\Omega$:

\begin{equation}
\Omega ( \rho \otimes |e_{(+)}\rangle\langle e_{(+)}| ) = \Lambda^{(+)} ( J^{-1} \rho ) \otimes |e_{(+)}\rangle\langle e_{(+)}| 
\end{equation}

\begin{equation}
\Omega ( \rho \otimes |e_{(-)}\rangle\langle e_{(-)}| ) = \Lambda^{(-)} ( K^{-1} \rho) \otimes |e_{(-)}\rangle\langle e_{(-)}| 
\end{equation}

$\Omega$ is completely positive since all its components, $\Lambda^{(+)}$, $\Lambda^{(-)}$, $J$ and $K$, are positive.  It is also trace preserving, since for the $(+)$ component:

\begin{equation}
Tr [ \Lambda^{(+)} ( J^{-1} \rho ) ] \\
= \Lambda^{(+)}_{r'r,r's} J^{-1}_{rt} \rho_{ts} \\
= J_{sr} J^{-1}_{rt} \rho_{ts} = Tr [\rho]
\end{equation}

For the $(-)$ component, singularities in $\Lambda^{(-)}$ poses a minor problem:

\begin{equation}
Tr [ \Lambda^{(-)} ( K^{-1} \rho ) ] = Tr [\Psi \rho]
\end{equation}

However, we note that after applying the extension map the domain of states is $S \rho$, and $\Psi (S \rho) = S \rho$, so this component map is trace preserving on this domain.

Therefore the map $\Omega$ is completely positive and trace preserving.  Putting $\Omega$ and $E$ together we have:

\begin{equation}
Tr_e[ \Omega ( E(\rho) ) ] = \Lambda (\rho)
\end{equation}

It is a known procedure to write the completely positive and trace preserving map $\Omega$ as a unitary transformation $u$ in an extended space.  The dimension of this space is $dim(J)*l_{(+)}+dim(K)*l_{(-)}$, where $l_{(+)}$ and $l_{(-)}$ are the number of eigenmatrices of $\Lambda^{(+)}$ and $\Lambda^{(-)}$ respectively, and $l_{(+)} + l_{(-)}\leq N^2$. 

\section{Conclusions}

We showed that any trace preserving map, whether positive or negative, can be expanded in terms of an extension map and a completely positive map.  We also described a procedure to make any map trace preserving.



\begin{thebibliography}{99}

\bibitem{Sudarshan61}E.C.G. Sudarshan, P.M. Mathews, J. Rau, {\it Stochastic 
Dynamics of Quantum-Mechanical Systems}, Phys. Rev. Vol.121 No.3, 920-924 
(1961)

\bibitem{Choi72}M. Choi, {\it Positive Linear Maps on C*-Algebras}, Can. J. 
Math Vol.24 No.3, 520-529 (1972)

\bibitem{purification}C.H. Bennett, G. Brassard, S. Popescu, B. Schumacher, J.A. Smolin and W.K. Wootters, {\it Purification of noisy entanglement and faithful teleportation via noisy channels}, Phys. Rev. Lett. 76, 722 (1996)

\bibitem{SudarshanShaji03}E.C.G. Sudarshan, A. Shaji, {\it Structure and Parametrization of Generic Stochastic Maps of Density Matrices}, J. Phys. A, 36, 5073-5081 (2003)

\bibitem{JordanShajiSudarshan04}T.F. Jordan, A. Shaji, E.C.G. Sudarshan, {\it Dynamics of initially entangled open quantum systems}, Phys. Rev. A 70, 1 (2004)

\end{thebibliography}
\end{document}